\documentstyle[aps]{revtex}

\begin{document}
\flushbottom
% run page numbers by "chapter"
\def\thepage{\roman{page}}
\title{\vspace*{1.5in}Peculiarities of anisotropy and polarization as 
an indicator of noises in the CMB maps}

\author{E. Kotok}

\address{Theoretical Astrophysics Center, Juliane Maries Vej
30, 2100 Copenhagen, {\O} Denmark\\ August 2000}

\author{P.Naselsky}

\address{Theoretical Astrophysics Center, Juliane Maries Vej
30, 2100 Copenhagen, {\O} Denmark;\\ Rostov State University, 
Zorge 5, 344090 Rostov-Don, Russia}

\author{D.Novikov}

\address{Astronomy Department, University of Oxford, NAPL, Keble 
               Road, Oxford OX1 3RH, UK;\\Astro-Space Center of 
               P.N.Lebedev Physical 
                Institute, Profsoyuznaya 84/32, Moscow, Russia}

\author{I.Novikov}

\address{Theoretical Astrophysics Center, Juliane Maries Vej
30, 2100 Copenhagen, {\O} Denmark;\\ Astronomical Observatory of the
Copenhagen University, Juliane Maries Vej 30,\\ 2100 Copenhagen, 
{\O} Denmark; \\ Astro-Space Center of Lebedev Physical Institute, 
Profsoyuznaya 84/32, Moscow, Russia;\\ NORDITA, Blegdamsvej 17, 
DK- 2100, Copenhagen, Denmark.}

\maketitle

\input{epsf}
\newcommand{\be}{\begin{equation}}
\newcommand{\ee}{\end{equation}} \newcommand{\g}{\nabla}
\newcommand{\de}{\partial}    \newcommand{\ha}{\frac{1}{2}}
\newcommand{\ci}[1]{\cite{#1}}  \newcommand{\bi}[1]{\bibitem{#1}}
\newcommand{\noi}{\noindent}

\newcommand{\ga}{\alpha}
\newcommand{\gb}{\beta}
\newcommand{\gc}{\gamma}
\newcommand{\gd}{\delta}
\newcommand{\gep}{\epsilon}
\newcommand{\gee}{\varepsilon}
\newcommand{\gz}{\zeta}
\newcommand{\get}{\eta}
\newcommand{\gth}{\theta}
\newcommand{\gthh}{\vartheta}
\newcommand{\gi}{\iota}
\newcommand{\gk}{\kappa}
\newcommand{\gl}{\lambda}
\newcommand{\gm}{\mu}
\newcommand{\gn}{\nu}
\newcommand{\gks}{\xi}
\newcommand{\go}{\0}
\newcommand{\gp}{\pi}
\newcommand{\gpp}{\varpi}
\newcommand{\gr}{\rho}
\newcommand{\grr}{\varrho}
\newcommand{\gs}{\sigma}
\newcommand{\gss}{\varsigma}
\newcommand{\gt}{\tau}
\newcommand{\gu}{\upsilon}
\newcommand{\gf}{\varphi}
\newcommand{\gff}{\varphi}
\newcommand{\gx}{\chi}
\newcommand{\gps}{\psi}
\newcommand{\gw}{\omega}
\newcommand{\gG}{\Gamma}
\newcommand{\gD}{\Delta}
\newcommand{\gTh}{\Theta}
\newcommand{\gL}{\Lambda}
\newcommand{\gKs}{\Xi}
\newcommand{\gP}{\Pi}
\newcommand{\gS}{\Sigma}
\newcommand{\gU}{\Upsilon}
\newcommand{\gF}{\phi}
\newcommand{\gPs}{\Psi}
\newcommand{\gW}{\Omega}

\newcommand{\ti}{\tilde}
\newcommand{\Li}{{\cal L}}
\newcommand{\ra}{\rightarrow}
\newcommand{\pa}{\partial}
\newcommand{\ov}{\overline}
\newcommand{\fad}{\frac{\Delta T}{T}}
\newcommand{\lan}{\langle}
\newcommand{\ran}{\rangle}
\flushbottom

\begin{abstract}
We discuss  some new problems of the modern cosmology which arose after 
the BOOMERANG and MAXIMA-1 successful missions. Statistics of high peaks 
of the CMB anisotropy is analyzed and we discuss possible inner structure 
of such peaks in  the observational data of future MAP and PLANCK 
missions. We have investigated geometrical and statistical properties of 
the CMB polarization around such high isolated peaks of anisotropy in the 
presence of a polarized pixel noise and point sources. The structure of 
polarization fields in the vicinity of singular points with zero 
polarization is very sensitive to the level of pixel noises and point 
sources in the CMB maps. 

PACS number(s): 98.80.Cq, 95.35.+d, 97.60.Lf, 98.70.Vc.
\end{abstract}

%\twocolumn
\section {Introduction}

Observational data by BOOMERANG and MAXIMA-1 [1,2] open a new epoch 
in the investigation of the CMB power spectrum at the large multipole 
numbers $l$. 
The measured angular power spectrum shows a clear peak at angular scales 
corresponding to the spherical harmonic multipole number $l\approx 200$. 
This is the Sakharov's peak. However the structure of the CMB anisotropy 
power spectrum at $l>400$ is still unclear and next generation of 
satellite experiments (such as MAP and PLANCK) are needed.

There exist a few factors which play a leading role in the future 
experiments. Namely, both the MAP and PLANCK missions will provide 
significantly greater percentage of sky coverage than BOOMERANG and
MAXIMA-1. In addition in the PLANCK mission there are two HFI channels
$\nu\simeq 545$GHz and $\nu \simeq 857$GHz which provide $FWHM=5$arcmin 
resolution. Future polarization measurements are also very important.

In contrast to BOOMERANG and MAXIMA-1 in both high frequency channels of 
PLANCK contamination foregrounds are of a prime consideration [3]. Of 
course, there is a wide variety of ``technical'' methods which could be 
applied to obtain a well-cleaned data over a wide range of angular scales.
For example, one can use the frequency dependence of the various 
foregrounds (dust emission, synchrotron, thermal emission, 
point sources etc.) [3,4]. In practice this method meets serious 
difficulties because of the angular distribution on the sky and the 
frequency dependence of the foreground components are not well known.

That is why we would like to focus our attention on the investigation of 
local characteristics of the CMB anisotropy and polarization. 

In this paper we analyze the statistics of the peaks (maxima and minima)
in the CMB anisotropy maps. We compare  this statistics for the
BOOMERANG and MAXIMA-1 maps with the statistics for the future MAP and 
PLANCK observations and predict some properties of the peaks and their 
shapes in the future PLANCK observations. The main point of our attention 
is the analysis of the structure of the polarization filed around singular
points with zero polarization. We propose to use the results of the 
analysis for the estimation of the level of noise and foregrounds in the 
maps.

\section{Statistics of peaks in the CMB anisotropy maps}

In this section we compare statistics of peaks (maxima and minima) for 
the BOOMERANG and MAXIMA-1 maps with the statistics for the future MAP 
and PLANCK observations. 

Let us consider the model which is close to the real situation provided by
the successful MAXIMA-1 balloon mission. According to [2] in the MAXIMA-1 
pixeled map one can find a high amplitude peak in $\gD T/T$ distribution 
on the observed region of the sky with following coordinates: declination
$\simeq 58.6$ degrees; RA$=15.35$ hours. In the Wiener-filtered map this 
peak has an amplitude  $\gD T\sim 2.3\div 2.5\gs$ and it decreases 
monotonically down to $1\gs$ cross-level at $15.2< h \le 15.4$ hours, 
$58.5< \gd <60$ degree. 

Below we predict some definite properties of the structure of such peaks 
in the future PLANCK CMB measurements performed with higher angular 
resolution than the current measurements. Analogous prediction has been 
done in [5] for the Tenerife Experiment using COBE DMR
data although with the different technique. For example, we show that more
accurate measurements will not reveal the inner structure of the peak in 
a form of new high $(>1.5\gs)$ peaks inside the area mentioned above. We
assume that the fluctuations of the CMB anisotropy (or polarization) are
realization of a random Gaussian process. It is well known that for such 
random processes all statistical properties of a signal are determined 
by the correlation function of the anisotropy
\be
C_T(\gth)=\frac{1}{4\pi}\sum_l(2l+1)C_{l,T}W(l)P_l(\cos \gth),
\label{cteq1}
\ee
where $C_{l,T}$ is the power spectrum of fluctuations of anisotropy, 
$W(l)$ is the window-function which depends on the strategy of the 
experiment, $P_l(\cos\gth)$ are Legendre polynomials. Analogous 
expressions can be written for $Q$ and $U$ components of polarization: 
$C_Q$ and $C_U$. 

To investigate topology of the anisotropy and polarization maps we define 
the spectral parameters (as in [6,7])
\be
\gs_i^{2}=(-1)^i i!2^{2i}\frac{d^iC(0)}{d\gw^i}; \hspace{0.5cm}
\gw=2\sin\frac{\gth}{2}; 
\label{gseq2}
\ee
\[
\gamma(\gth_A)=\frac{[\gs_1]^2}{\gs_0\gs_2};
\hspace{0.5cm} \gth_*=\sqrt{2}\frac{\gs_1}{\gs_2};
\hspace{0.5cm} (\gth_C)^2=-C(0)/\frac{d^2C(0)}{d\gth^2},
\]
where $i=0,1,2$, and $\gth_A$ is the antenna beam.

In the real experiments the antenna has a finite resolution and the 
spectral parameters depend on the antenna beam $\gth_A$ and therefore on 
the number of Doppler peaks which could be resolved by the antenna. This 
means that in the future PLANCK maps the structure of the  high $\gD T/T$ 
peaks can be different from the corresponding structure in the BOOMERANG 
and MAXIMA-1 maps.

For all cosmological models the power spectrum of the CMB anisotropy 
$C_l$ can be described in terms of a sum of the Gaussian peaks centered at
the points of the  maxima $l_n$, $(l\ge 30)$ ([8])\footnote{We will omit 
the index $T$ in the subsequent discussions}
\be
\frac{l(l+1)C_l}{2N\pi}=
\left\{\sum_{n}A_n \exp{\left[-\frac{(l-l_n)^2}{2d_n^2}\right]+1}\right\}
e^{-l^2s^2},
\label{eq3}
\ee
where $n$ is the number of the peak, $d_n$ is the width of the peak, 
$l_n$ is its position, $A_n$ is its amplitude and $N$ is the 
normalization factor for low multipole range (for example, the COBE data 
normalization). The last term in Eq.(\ref{eq3}) accounts for the 
Silk-damping at the angular scale $s$. Note that we do not include in 
Eq.(\ref{eq3}) any low and high multipole filters (the window function and
beam). These means that Eq.(\ref{eq3}) describes the initial power 
spectrum of fluctuations on the sky without any smoothing. In reality both
last factors are extremely important and their influence on the $\gD T/T$
maps plays a crucial role. The power spectrum in the form of 
Eq.(\ref{eq3}) after substitution into Eq.(\ref{gseq2}) gives information 
about the influence of each peak on the topology of $\gD T/T$ maps. Thus, 
using the approximate Eq.(\ref{eq3}) we can investigate the influence of 
the first, second and subsequent Doppler peaks on the spectral parameters 
of the future maps of the MAP, PLANCK and other missions.

Next point is connected with the window function $W(l)$ of the 
experiments. In a small angular approximation we will model general 
properties of $W(l)$ as follows:
\be
G(l)=\frac{W(l)}{l}=\exp{[-l(l+1)\gth_A^2]}\left\{  \begin{array}{ll}
l^m & \mbox {if $l  \ll 30$}\\
l^{-1} & \mbox {if $l \gg 30$},
\end{array}
\right. 
\label{eq4}
\ee
where $m=2$ for the single difference and $m=3$ for the double difference 
scheme of the low multipole filtration.
The exponent in Eq.(\ref{eq4}) describes the antenna beam with 
$\gth_A\simeq 7.45\times 10^{-3}\left(\frac{\gth_{FWHM}}{1^o}\right)$. 
For description of the asymptotes of the multiplier in Eq.(\ref{eq4}) we 
can introduce a function which matches both limits
\be
g(l)\simeq \frac{(lR)^{m+1}}{l[1+(lR)^{m+1}]}\hspace{0.2cm},
\label{gleq5}
\ee
where $R\sim 0.03$ is the characteristic angular scale at the low 
multipole filtration [see Eq.(\ref{eq4})]. So, in such a model the 
spectral parameters in Eq.(\ref{gseq2}) are:
\be
\gs^2_i=\int^{\infty}_0 dl l^{2i}g(l)\left[1+\sum_nA_n
\exp\left(-\frac{(l-l_n)^2}{2d_n^2}\right)\right]e^{-l^2(s^2+\gth^2_A)}; 
\hspace{0.5cm}i=0,1,2.
\label{gseq6}
\ee
Note that for the second and higher Doppler peaks we have 
$l_n^2/d_n^2\gg 1$
and only for the first Doppler peak $l_1^2/d_1^2\simeq 5$. For the 
analytical approximation of the integral in Eq.(\ref{gseq6}) we require 
the asymptotic $l_n^2/d_n^2\gg 1$ for all peaks in the power spectrum 
Eq.(\ref{eq3}). Using this approximation we obtain the following result 
for the spectral parameters $\gs^2_i$:
\be
\gs^2_0 = 
\ha\left\{2\ln\frac{R}{\xi}-\mbox{\boldmath$C$}+\sqrt{\frac{\pi}{2}}
\sum_n A_n\frac{d_n}{l_n}\exp{\left(\frac{-l_n^2\xi^2}{1+2d_n^2\xi^2}
\right)}\cdot(1+2d_n^2\xi^2)^{1/2}\right\},
\label{eq7}
\ee
\be
\gs^2_1 = \frac{1}{2\xi^2}+\sqrt{\frac{\pi}{2}}\sum_n
\frac{A_nl_nd_n \exp\left(\frac{-l_n^2\xi^2}{1+2d_n^2\xi^2}\right)}
{(1+2d_n^2\xi^2)^{3/2}}
\left[1+\Phi\left(\frac{l_n}{d_n\sqrt{2(1+2d_n^2\xi^2)}}\right)\right],
\label{eq8}
\ee
\be
\gs^2_2 = \frac{1}{2\xi^4}+\sqrt{\frac{\pi}{2}}\sum_n
\frac{ A_nl^3_nd_n \exp\left(\frac{-l_n^2\xi^2}{1+2d_n^2\xi^2}\right)}
{(1+2d_n^2\xi^2)^{7/2}}
\left[1+\Phi\left(\frac{l_n}{d_n\sqrt{2(1+2d_n^2\xi^2)}}\right)\right],
\label{eq9}
\ee
where $\mbox{\boldmath$C$}$ is the Euler constant, $\xi^2=\gth_A^2+s^2$, 
$\Phi(x)=2/\sqrt{\pi}\int_0^x dx e^{-x^2}$ is the probability integral. 
As we can see from Eq.(\ref{eq7}) only the first Doppler peak is important
for calculation of the variance $\gs^2_0$. The influence of the second
and higher peaks is practically negligible due to the decrease of 
amplitudes $A_n$ and $d_n/l_n$. However these peaks determine 
(see Eqs.(10)-(12)) the topological structure of $\gD T/T$ maps (for 
example, the number of maxima and minima at different thresholds 
$\nu_n\gs_0= \gD T/T$). Using Eqs.(8) and (9) we describe the realistic 
model  at $d_n^2\xi^2\ll 1$ and $l_n^2\xi^2\le 1$. In such a model the 
density of all peaks at $\nu\in(-\infty,\infty)$ has an especially simple 
form
\be
N^+_{PK}=N^-_{PK}=\frac{1}{8\pi\sqrt{3}}\frac{\gs^2_2}{\gs^2_1}
\hspace{0.2cm}{\mbox{(ster)}}^{-1},
\label{npkeq10}
\ee
where $N^+_{PK}$ and $N^-_{PK}$ are the densities of all maxima and minima
correspondingly. The density of saddle points (of arbitrary hight) is
\be
N_{sad}(-\infty)=
2N^+_{PK}\hspace{0.2cm}\hspace{0.2cm}{\mbox{(ster)}}^{-1}.\label{nsadeq11}
\ee
Let us now discuss a model in which all Doppler peaks are smoothed 
$(A_n=0)$. In such a model the spectral parameters $\gth_*$ and $\gamma$ 
are the following
\be
\gth_*^2=2\xi^2;\hspace{1cm}
\gamma=(2\ln\frac{R}{\xi}-\mbox{\boldmath$C$})^{-1/2},
\label{gtheq12}
\ee
and the densities of all maxima and minima (with arbitrary hight) are
$N^+_{PK}=N^-_{PK}=\frac{1}{8\pi\sqrt{3}}\xi^{-2}$. If the sky 
coverage for some $\gD T/T$ - experiment is $f_{sky}$ (for example, 
$f_{sky}\simeq0.3\%$ for the MAXIMA-1 experiment), then the number of 
maxima (or minima) in the observational map is
\be
N_{max}\simeq 16\left(\frac{f_{sky}}{0.003}\right)
\left(\frac{\gth_{FWHM}}{1^o}\right)^{-2}.
\label{nmaxeq13}
\ee

According to [1,2] for MAXIMA-1 and BOOMERANG  the antenna beam 
corresponds to $FWHM\simeq 10'$. This means that in the absence of the 
Doppler peaks in the power spectrum we can find 576 maxima on the 
corresponding maps. However one can find from Eq.(7)-(9) that the presence
of Doppler peaks in the initial power spectrum changes the number of peaks
in the map down to 271 for the mentioned above observational data. So, 
after our analysis we can conclude that the influence of the Doppler peaks
 leads to decrease of the number of hot and cold spots on the map by a 
factor $\simeq 2$. This result consists with the data of the MAXIMA-1 and 
BOOMERANG maps.

The next problem which we would like to discuss below is the following.
How sensitive is the topology of $\gD T/T$ map to the amplitudes of the 
second $A_2$ and third $A_3$ Doppler peaks assuming that the amplitude and
 the position of the first one are known? To give answer to this question
we compare $\gamma(A_2,A_3)$ and $N_{pk}(A_2,A_3)$ for the following 
models. In the first model we take the amplitude $A_1$ corresponds to the 
data [1,2] and positions and widths of the next peaks in power spectrum 
are the following: $l_1= 210$ with the width $d_1= 95$, $l_2= 580$ with 
the width $d_2= 110$ and $l_3= 950$ with the width $d_3= 130$. 
Corresponding plots are in Fig.1.
%%----------------------------Figure 1----------------------

\begin{figure}[h]
\vspace{0.01cm}\hspace{-0.1cm}\epsfxsize=7cm 
\epsfbox{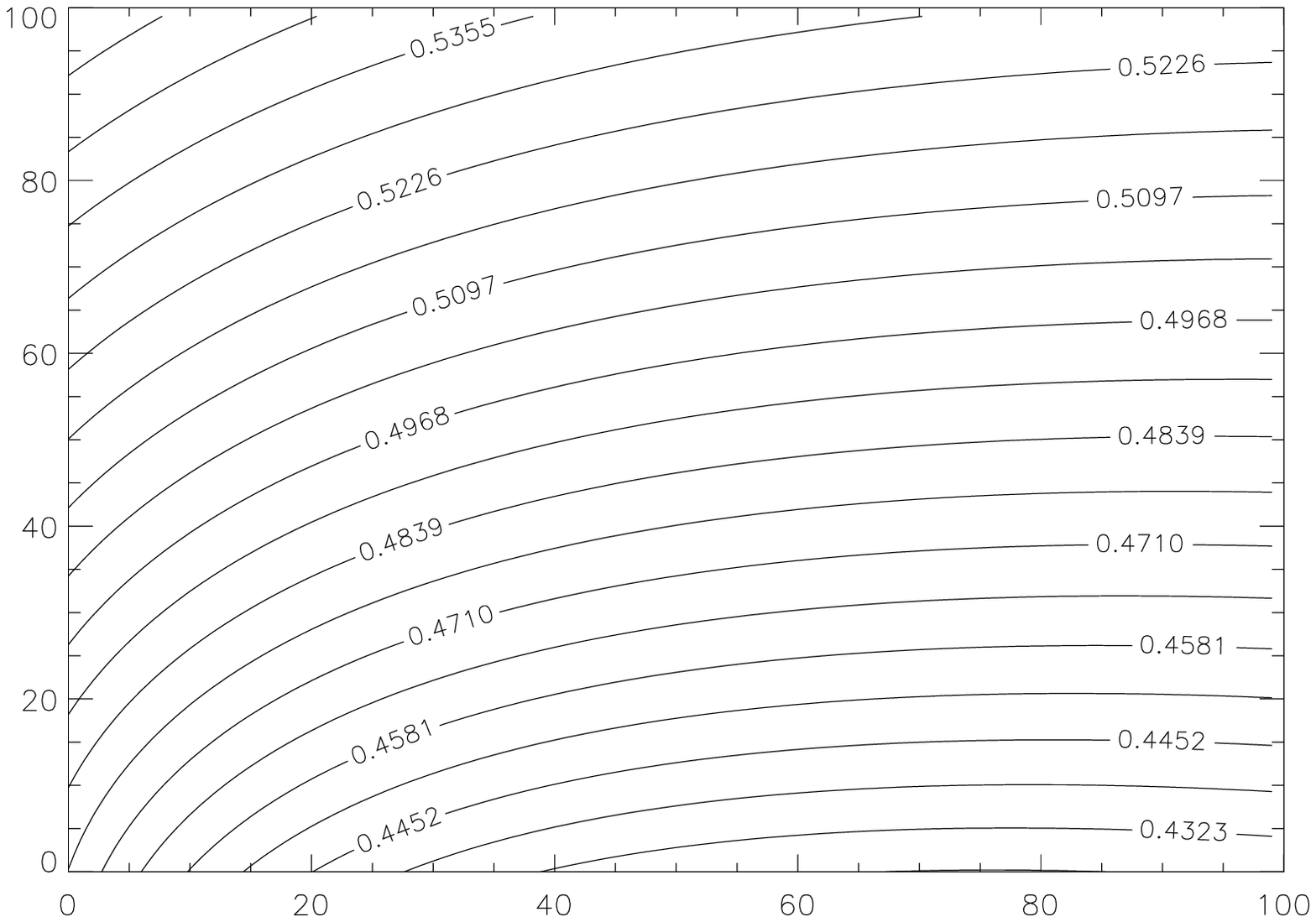}
\hspace{0.0cm} \epsfxsize=7cm 
\epsfbox{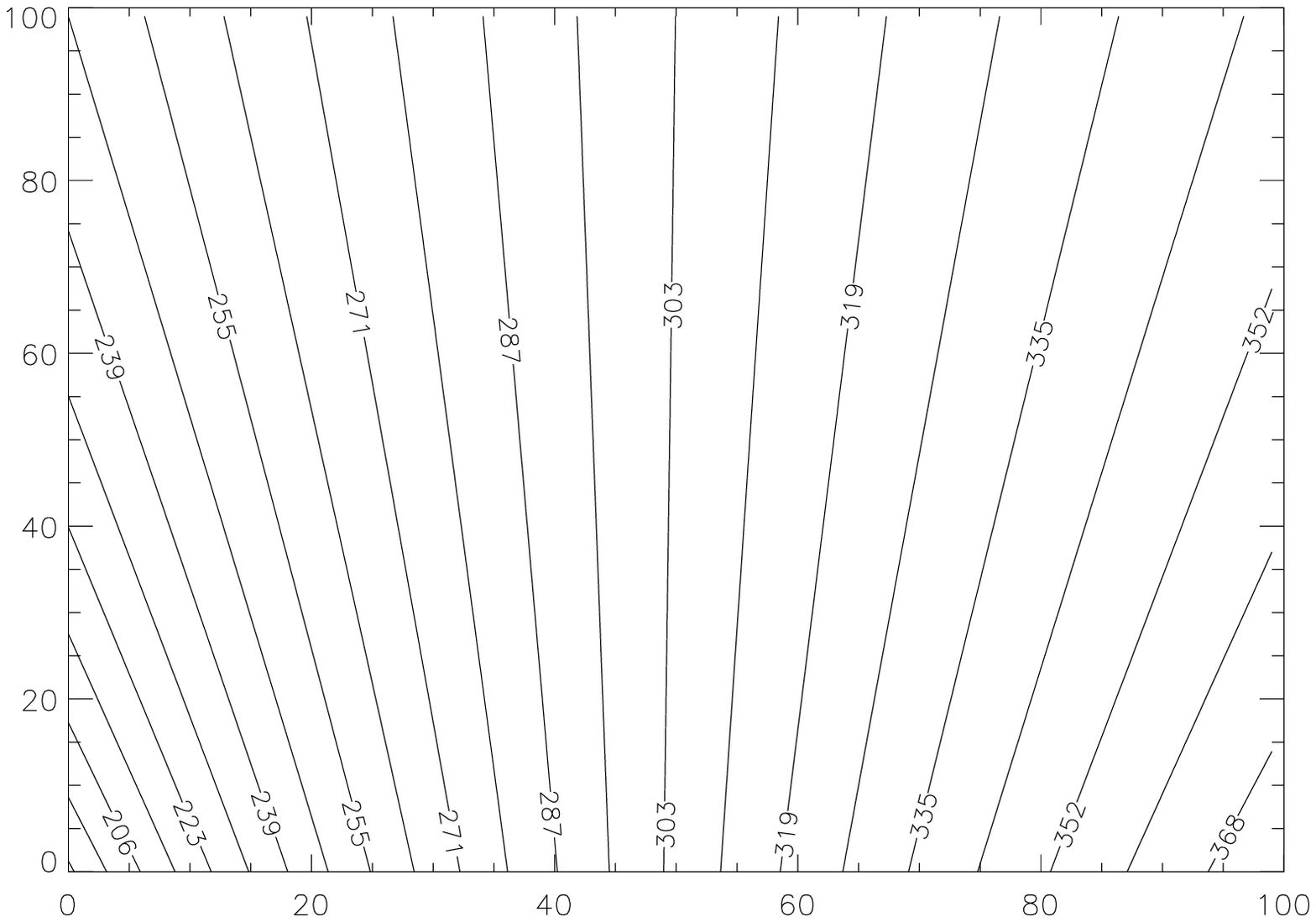} 
    \caption{The dependence of the $\gamma(x,y)$ (left) and $N_{pk}(x,y)$ 
             (right) on the parameters $x=10^2A_2/A_1$ (horizontal axes), 
             $y=10^2A_3/A_1$ (vertical axes). The numbers in the lines 
             correspond to the values of $\gamma(x,y)$ and $N_{pk}(x,y)$.
             Note that $x=26$ and $y=46$ corresponds to the MAXIMA-1 and 
             BOOMERANG amplitude of the first, and two next Doppler 
             peaks.}
\end{figure}
%---------------------------Figure 1------------- 

The second model (see plots in Fig.2) corresponds to a hypothetical 
situation when the amplitude of the first Doppler peak is two times 
smaller than in the previous case. As one can see from Eq(7)-(9) and 
Fig.1, Fig.2, in this second ``toy'' model the structure of the spectral 
parameters $\gamma(A_2,A_3)$ and $N_{pk}(A_2,A_3)$ changes drastically. 
The number 
of the maxima  increases up to more than 420 while $\gamma$ parameter 
conserves practically the same value: $\gamma\simeq 0.4-0.47$. This result
is important for analysis of the global and local topology of the maps. 
In the BOOMERANG and MAXIMA-1 experiments the amplitude and position of 
the first Doppler peak in $C_l$-power spectrum are measured with 10\% 
accuracy. That means that theoretical predictions of the number of the CMB
peaks in the observational maps could change from 263 to 279 due to this 
10\% uncertainty. The difference in 16 peaks corresponds to 
$\frac{\gd N}{N}\sim N^{-1/2}$ statistical fluctuations of the number of 
peaks $N$ in the map practically without any essential changes of the 
$\gamma$-parameter. For other aspects of the distribution of the peaks in 
the maps see [7-9].
%----------------------------Figure 2----------------------
\begin{figure}[h]
\vspace{0.01cm}\hspace{-0.1cm}\epsfxsize=7cm 
\epsfbox{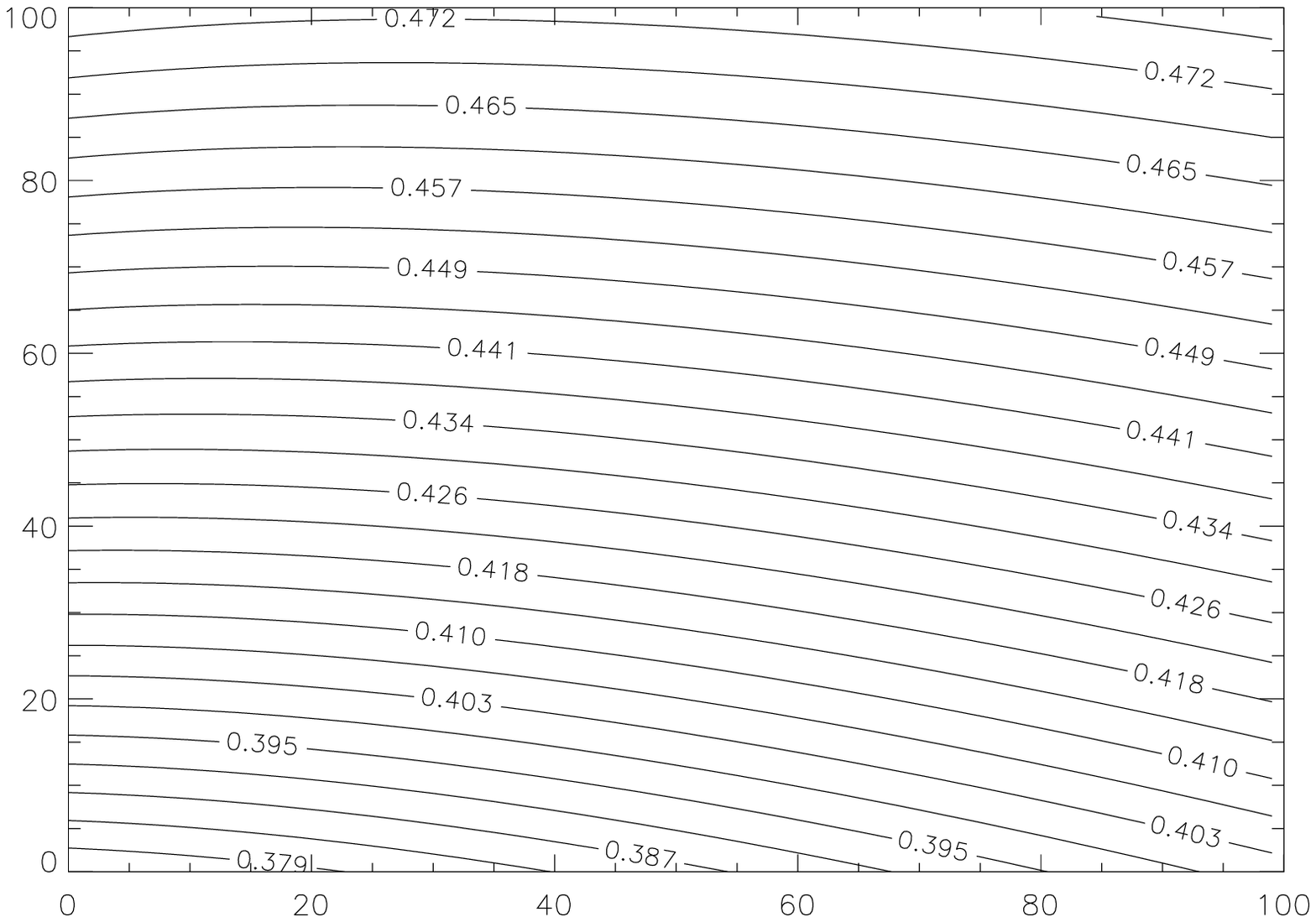}
\hspace{0.0cm} \epsfxsize=7cm 
\epsfbox{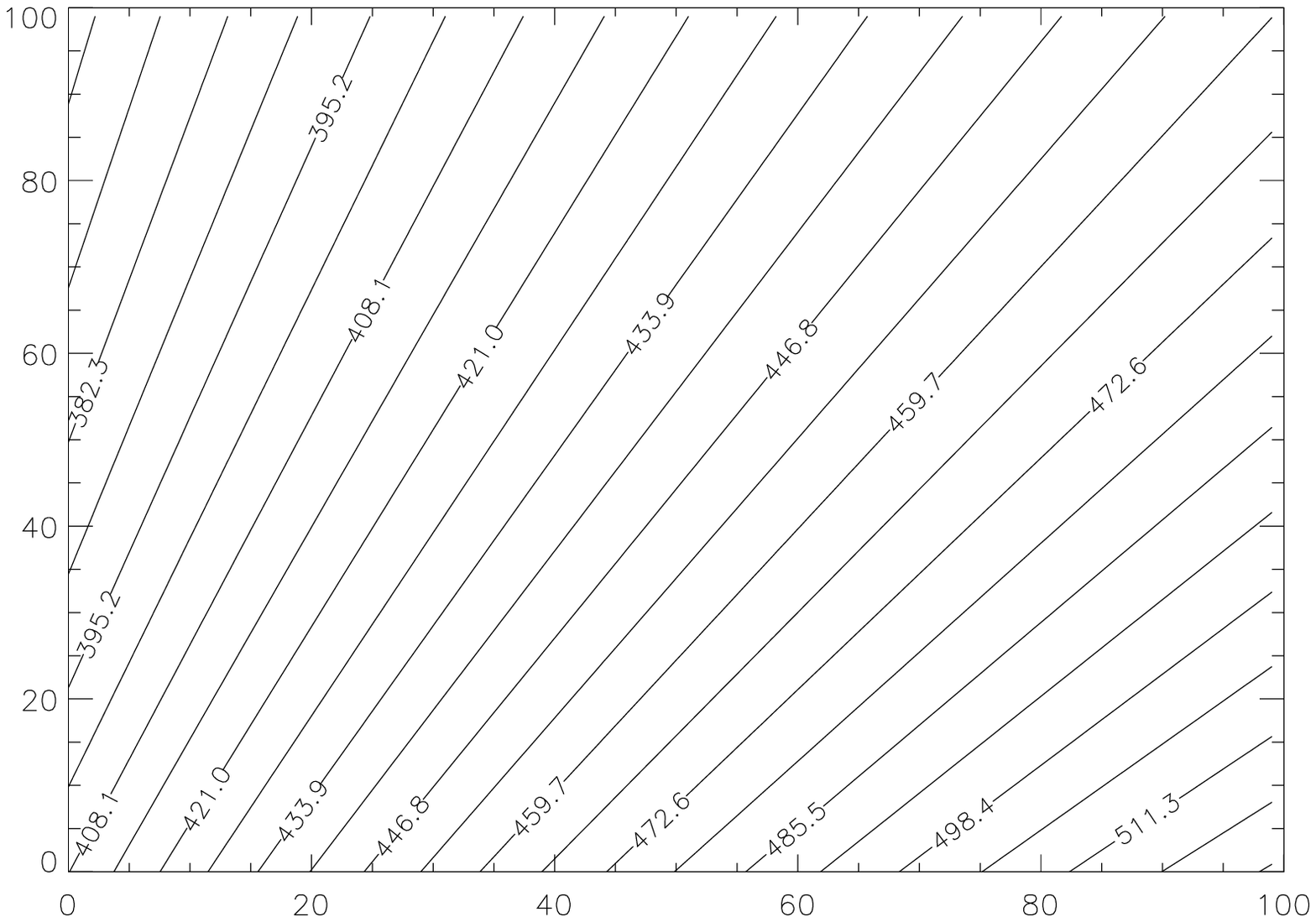} 
     \caption{The same as in Fig.1 but for a toy model with the 
              amplitude of the first Doppler peak is two times 
              less than in Fig.1 }
\end{figure}
%---------------------------Figure 2-------------
\section{Structure of the high peaks of the CMB anisotropy.} 

In this section we consider how the structure and shape of the high peaks 
in the observational maps change when one goes to higher angular resolution 
maps. In this case one can measure the structure of the $\gD T$-field 
around a high peak with more details.

Can the future measurements reveal internal structure of the peaks which are
found by BOOMERANG and MAXIMA-1?. For example can the future measurements 
reveal new peaks in the fine structure inside the region around the 
$\nu\simeq 2$ - peak down to the $\nu=0$? And if yes, what would be a 
typical height of such peaks? The answers to these questions depend on the 
peak-peak correlation in a high-resolution map. Two important properties of 
a random Gaussian field play a leading role in this problem. First, in the
vicinity of a high maximum of a Gaussian field the random character of 
$\gD T(x,y)$ distribution is broken, and the shape of the peak is regular. 
The typical scale of such a regularity is, of course, $\gth_*$ 
(see Eq.(\ref{gseq2})). Second, as we mentioned above, the fractional 
variance in the peak number from one realization to another over a grid of 
area $\gW_p$ is related to the peak-peak correlation function $C_{pk-pk}$ 
[6,9]:
\be
\lan(\gD N^+_{pk})^2\ran/\lan  N^+_{pk}\ran^2=
\lan N^+_{pk}\ran^{-1}+
\int\frac{d\gW_{\ov{q}}d\gW_{\ov{q}'}}{\gW^2_p}C_{pk-pk}(\ov{q}-\ov{q}'),
\label{eq14}
\ee
where $\lan N^+_{pk}\ran= n^+_{pk}(\nu_t)\gW_p$; $n^+_{pk}(\nu_t)$ is the 
integrated number density of the maxima with height $\nu$ above some 
threshold $\nu_t$. 
Note that the first term in Eq.(\ref{eq14}) corresponds to the Poissonian 
distribution of peaks. Recently Heavens and Sheth [9] performed analytical 
and numerical calculations of the  peak-peak correlation function and 
showed that $C_{pk-pk}$ goes to zero at $\gth<\gth_*$, and reaches negative 
value $C_{pk-pk}=-1$, at $\gth=0$. This result  reflects the fact that  the 
different high peaks cannot be located close to each other and, for example,
two high peaks with the amplitudes $\nu_1\sim\nu_2\sim 2\div2.5\gs$ should 
be separated by the distance $\gth\gg\gth_*$. According to [9] the typical 
angular scale $\gth_*$ for the most preferable $\gL$CDM cosmological models 
is close to 20 arcmin. This scale is twice greater than FWHM in the 
BOOMERANG and MAXIMA-1 experiments and 4 times greater than in the 
PLANCK-mission. However, it is worth noting, that inside the mentioned 
above region around the high peak there can exist up to $\sim 10$ low 
amplitude peaks ($\nu\le 1$) in  a high resolution map. These peaks are 
extremely important for the proof of the Gaussian statistics of a signal in 
such a map.  Thus we can conclude that the isolated $2\div 2.5\gs$ peaks 
which were found in the low-resolution BOOMERANG and MAXIMA-1 maps  will 
reveal themselves as isolated peaks in the PLANCK map.

Let us come back to the discussion of the  high peak at the location
$\gd=58.6$ degrees; RA$=15.35$ hours of the MAXIMA-1 map. The position of 
this peak practically does not depend on the higher angular resolution of 
the future PLANCK mission and its amplitude can be described as follows. 
Let us imagine that an ideal experiment with a $\gd$-function antenna beam 
finds the highest peak in the $\gD T$ map at the position 
$(\ov{\gd};\ov{RA})$. An amplitude of such a peak measured in units  
of variance is:
\be
\nu_{in}= \gD T/\gs_{0(in)},
\label{eq15}
\ee
where $\gs_{0(in)}$ corresponds to Eq.(\ref{eq8}) at $\gth_A=0$ and $\xi=s$.
We suppose for simplicity that the distribution of $\gD T(x,y)$ around 
the point of maximum is Gaussian (more comprehensive consideration can be 
found in [6]) with the characteristic scales $a$ 
and $b=\gk a$, $\gk$ is a constant:
\be
\gD T(x,y)=
\nu_{in}\gs_{0(in)}\exp\left(-\frac{x^2}{2a^2}-\frac{y^2}{2b^2}\right),
\label{eq16}
\ee
where the parameter $a$ is proportional to the typical correlation scale of 
the initial signal. Following [6] we can describe a local shape of the peak
of height $\nu$, measuring radial curvature $\Gamma$ and ``ellipticity''$e$
using polar coordinates $\ov{\gth}$ and $\ov{\gff}$:
\be
\gd(\ov{\gth},\ov{\gff})=\gs_{0(in)}\left[\nu_{in}-\ha\gamma \Gamma\left(
\frac{\ov{\gth}}{\gth_C}\right)^2(1+2e \cos 2\ov{\gff})\right].
\label{eq17}
\ee
Let us denote $\ov{\gth}^2=x^2+y^2$ and 
$\cos2\ov{\gff}=\frac{x^2-y^2}{x^2+y^2}$. After that we find:
\be
a^2=\frac{\nu_{in}\gth_c^2}{(1+2e)\gamma\Gamma};\hspace{0.7cm}
b^2=\frac{\nu_{in}\gth_c^2}{(1-2e)\gamma\Gamma};\hspace{0.7cm}
\gk^2=\frac{1+2e}{1-2e}.
\label{eq18}
\ee
In terms of eccentricity $\gep=2\left[e/(1+2e)\right]^{1/2}$ the last 
relation in Eq.(\ref{eq18}) can be transformed to
\be
\gk^2=1/(1-\gep^2).
\label{eq19}
\ee
Let us consider two experiments which measure the same part of the sky 
around a peak with different angular resolutions $\gth_1$ and $\gth_2$.
Suppose that $\gth_1$ corresponds to the MAXIMA-1 FWHM and $\gth_2$ 
corresponds to the PLANCK $(\gth_1\simeq 2\gth_2$). We denote an amplitude 
of the maximum in the low-resolution experiment by $\nu_{maxima}$ and for 
the high-resolution experiment by $\nu_{planck}$. 
For such models an amplitude of the peak is
\be
\widetilde{\gD T}_j(x,y)=
\frac{1}{2\pi\gth^2_j}\int dx'dy'\gD T(x',y')
\exp\left[-\frac{(\ov{r}-\ov{r}')^2}{2\gth^2_j}\right],
\label{DTeq20}
\ee
where index $j=1,2$ corresponds to $\gth_1$ and $\gth_2$, 
$\ov{r}(x,y)$ and $\ov{r}'(x',y')$ are vectors in the 
Cartesian coordinate system centered at the point of the maximum.
The shape of $\widetilde{\gD T}_j(x,y)$ from Eq.(\ref{eq16}) for
the low- and high-resolution experiments is
\be
\widetilde{\gD T}_j(x,y)=
\frac{\nu_{in}\gs_{0(in)}ab}{[(a^2+\gth_j^2)(b^2+\gth_j^2)]^{1/2}}
\exp\left[-\frac{x^2}{2(a^2+\gth^2_j)}-\frac{y^2}{2(b^2+\gth^2_j)}\right].
\label{DTeq21}
\ee
This curve determines parameter 
$\xi^2_1=(b^2+\gth_1^2)/(a^2+\gth_1^2)$ which we can measure in the 
vicinity of the peak at some threshold $\nu_t\gs_0^{(1)}$, where 
$\gs_0^{(1)}$  is the variance of fluctuations in the low-resolution 
experiment. As a result the amplitude of the peak from Eq.(15) 
is
\be
\nu_{maxima}\gs^{(1)}_0=
\frac{\nu_{in}\gs^{(in)}_0\gk}{\xi_1(1+\gth^2_1/a^2)}.
\label{numaxeq22}
\ee
For the high-resolution experiment we obtain from Eq.(15) 
\be
\nu_{planck}\gs^{(2)}_0\simeq
\frac{\nu_{in}\gs^{(in)}_0\gk}{[(1+\gth^2_2/a^2)
(\gk^2+\gth^2_2/a^2)]^{1/2}},
\label{eq23}
\ee 
and while the difference between $\gs^{(in)}$, $\gs^{(1)}_0$ and
$\gs^{(2)}_0$ is logarithmic only (see Eq.(\ref{eq8})) we have
\be
\nu_{planck}\approx\nu_{maxima}\frac{\xi_1(1+4\mu^2)}
{[(1+\mu^2)(\gk^2+\mu^2]^{1/2}},
\label{eq24}
\ee
where $\mu=\frac{\gth_2}{a}<1$. 
 
For example, for the MAXIMA-1 peak at the coordinates $\gd=58.6$ degrees, 
RA$\simeq15.35$ this ratio is $\sim 1.2\div 1.4$. Taking into account 
this result %(see Eq.(\ref{eq8})), we can conclude that 
$\nu_1\simeq 2\div3$ 
peaks in the MAXIMA-1 map will be transformed to higher peaks in the 
PLANCK map ($\nu_2\simeq 3\div4$), $\gth_1\simeq 2\gth_2$. This 
conclusions are important for our discussions in the next section.

\section {Structure of the polarization field around the points where 
polarization of the CMB signal vanishes.} 

The structure of the polarization field and its auto- and 
cross-correlation with the anisotropy distribution on the sky has been 
studied in many papers (see for review [10]). Below we consider the 
structure of the polarization field around such singular points where 
$P=0$ in the CMB signal taking into account possible foregrounds and 
noise. This analysis can help in the estimation of a possible level of 
the foregrounds and noise in the  future MAP and PLANCK polarization maps.

Of course, measurements of the CMB polarization around the points where 
it vanishes are  a very difficult observational problem. However it 
is not necessary to go exactly to the point where $Q^2+U^2=0$ (where $Q$ 
and $U$ are Stocks parameters), because the structure of the polarization 
field can be determined by the pattern created by the flux lines around 
peculiar points. We note also that in spite of the fact that these points 
are distributed almost randomly over a map it is more probably to find 
them in the vicinities of high peaks of the $\gD T/T$ anisotropy. Indeed, 
in coming paper we will show that there is non-zero correlation between 
high maxima of anisotropy and singular points of polarization field.
In this paper we consider the problem in the continues (non discrete)
measurements of the $Q$ and $U$ values to reveal the main qualitative 
behaviors of the polarization field. The correspondingly consideration for 
pixelized maps see in [11].
We shall assume that in the observational map $Q$ and $U$ components of 
Stokes vector are the sum of the initial CMB signal $(\ov{Q},\ov{U})$ 
and the Gaussian noise $(n_Q,n_U)$ which model the influence of pixel 
noise and polarized point sources on the primordial CMB signal. We shall 
assume also that statistical properties of the noises $n_Q$ and $n_U$ are 
identical and their correlation functions are 
$C_{ij}^Q=C_{ij}^U=\gs^2_n\gd_{ij}$, where $\gs^2_n$ is the variance. 
This means that we neglected the correlation between noise of different 
pixels. Under the assumption  mentioned above we can expand $\ov{Q}$ and 
$\ov{U}$ near the points where $\ov{Q}=\ov{U}=0$ as follows:
\be
\begin{array}{l}
\ov{Q}\approx q_1x+q_2y\\
\ov{U}\approx u_1x+u_2y,
\end{array}
\label{qu}
\ee
where $x$ and $y$ are Cartesian coordinates of the system centered at the 
zero-point of polarization; $q_1$, $q_2$ and $u_1$, $u_2$ are the 
corresponding partial derivatives.
Thus the modulus of the polarization vector $P^2=Q^2+U^2$ is
\be
P^2=\ov{P}^2+2(\ov{Q}n_Q+\ov{U}n_U)+n_Q^2+n_U^2,
\label{p2}
\ee
where $\ov{Q}$ and $\ov{U}$ correspond to Eq.(\ref{qu}).\\
Lets us firstly describe the model when $n_Q$ and $n_U$ are negligible 
with respect to the regular functions $\ov{Q}$ and $\ov{U}$. This 
situation takes place when $\ov{Q}^2+\ov{U}^2\gg n_Q^2+n_U^2$ or 
$\ov{P}^2\gg\gs_n^2$, where $\gs^2_n$ is variance of the noise and 
in Eq.(\ref{p2}) we can neglect all terms excluding $\ov{P}^2$. In such 
a case we obtain from Eq.(\ref{p2}):
\be
P^2=\ov{P}^2=\sum_{i,j=1}^2 a_{ij}x^iy^j,
\label{p22}
\ee
where $a_{11}=q_1^2u_1^2$; $a_{22}=q_2^2u_2^2$; 
$a_{12}=a_{21}=q_1q_2+u_1u_2$. Rotating the coordinate system 
through the  
angle tg$2\psi=\frac{2a_{12}}{a_{11}-a{22}}$  one can transform 
Eq.(\ref{p22}) to the ``standard'' form:
\be
\ov{P}^2=Ax^2+By^2,
\label{ovp}
\ee
where $A=\ha\left[a_{11}+a_{22}+\sqrt{1+t^2}(a_{11}-a_{22})\right]$;
$B=\ha\left[a_{11}+a_{22}-\sqrt{1+t^2}(a_{11}-a_{22})\right]$; 
$A\cdot B>0$; $t=$tg$2\psi$ and $x$ and $y$ are the coordinates after 
rotation. 

As one can see from Eq.(\ref{ovp}) the condition $\ov{P}=const$
determines the ellipses with a peculiar isolated point inside them 
$\ov{P}^2=0$. Therefore the structure of the field $P^2(x,y)$ around 
points $\ov{P}^2=0$ for the case $\ov{P}^2\gg n_Q^2+n_U^2$ is rather 
simple. But when we come close enough to 
the point $\ov{P}^2=0$ this condition will be violated because the level 
of noise is approximately constant $>0$ near the point $\ov{P}^2=0$. Now 
in Eq.(\ref{p2}) we take into account the second term but neglect 
$n_Q^2+n_U^2$, so
\be
P^2-\ov{P^2}=2(\ov{Q}n_Q+\ov{U}n_U).
\label{pp}
\ee

As one can see from Eq.(\ref{pp}) the function $f=P^2-\ov{P^2}$ is a 
nonuniform random function due to a regular character of $\ov{Q}$ and 
$\ov{U}$ components in the vicinity of the peculiar point and the 
randomly distributed noise. From Eq.(\ref{pp}) we can estimate the 
variance of the random function $f$:
\be
\gs_f^2\approx 4\ov{P}^2\gs^2_n,
\label{sigma}
\ee
which depends on the module $\ov{P}(x,y)$.  From Eq.(\ref{sigma}) we can
define more precisely the limits of applicability of the approximation 
Eq.(\ref{pp}): $\gs_f^2 > \gs^4_n$ and therefore 
\be
\gs_n < \ha\ov {P}.
\label{gsn}
\ee
This relation shows that 
influence of the noise on the contour lines around peculiar point of 
polarization field becomes stronger when we come closer to the point 
$\ov{P}=0$. 
 
Finally at the very vicinity  of the point $\ov{P}^2=0$ we come to 
the region where $\gs^2_n > \frac{1}{4}\ov{P}^2$ and for this region we 
have the approximation
\be
P^2=n_Q^2+n_U^2.
\label{pnn}
\ee                                               
If $n_Q$ and $n_U$ are independent random Gaussian functions then 
$P=\sqrt{n_Q^2+n_U^2}$ is the Rayleigh random process with pdf $\Phi(U)$:
\be
\Phi(U)=\frac{U}{\gs_n^2}\exp\left(-\frac{U^2}{2\gs^2_n}\right).
\label{last}
\ee
As we can see from Eq.(\ref{last}) the probability to find the point with 
zero level of polarization inside the area 
$\gs_n^2 > \frac{1}{4}\ov{P}^2(x,y)$ is proportional to 
$\ov{P}^2/(4\gs_n^2)$ and goes to zero if $\ov{P}\ra 0$. This means that 
pixel noise or (and) polarized point sources  destroy the structure of 
the field around nonpolarized points of the primordial signal and in 
principle remove this peculiarities from the map. Thus the regions 
around the peculiar points of polarization are very sensitive to the pixel
and point sources noises. The structure of the polarization filed around 
such points gives us the information about the variance and statistical 
properties of the noises and can be used for filtration of such noises. 
We discuss this problem in details in [11] for filtration of the 
polarization maps.

\section{Conclusions} 
New theoretical methods have been developed recently to provide the 
analysis and exploiting the new and forcoming CMB data sets.
The BOOMERANG and MAXIMA-1 missions have discovered a lot of ``hot'' 
spots of the CMB radiation on the sky. These peaks in 
$\gD T$-distribution on the sky play a role of ``standard'' sources
 of the CMB anisotropy signal for future balloon and satellite 
experiments with more sensitive radiometers and higher angular 
resolutions. The new theoretical methods are developing now to provide 
the analysis and predictions the new and forthcoming CMB data sets.

\section{ Acknowledgments}
D.Novikov is grateful to the staff of TAC for providing excellent 
working conditions during his visit. 
The authors are grateful to Per Rex Christensen and Andrei Doroshkevich 
for useful discussions.
This investigation was supported in part by the Danish Natural 
Science Research Council
through grant No. 9701841 and also in part by the grants INTAS- 1192 
and RFFI- 17625. Authors are grateful to Danmarks Grundforskningsfond 
through its 
support for establishment of the Theoretical Astrophysics Center. 
D.Novikov acknowledges Board of the Glasstone Benefaction.

\end{document}